\documentclass[a4paper]{article}

\usepackage{INTERSPEECH2021}
\usepackage{multirow}
\usepackage[table,xcdraw]{xcolor}

\title{Additive Phoneme-aware Margin Softmax Loss for Language Recognition}
\name{Zheng Li$^1$, Yan Liu$^1$, Lin Li$^1$, Qingyang Hong$^2$}
\address{
	$^1$School of Electronic Science and Engineering, Xiamen University, China\\
	$^2$School of Informatics, Xiamen University, China}
\email{lilin@xmu.edu.cn, qyhong@xmu.edu.cn}

\begin{document}

\maketitle
\begin{abstract}
This paper proposes an additive phoneme-aware margin softmax (APM-Softmax) loss to train the multi-task learning network with phonetic information for language recognition.
In additive margin softmax (AM-Softmax) loss, the margin is set as a constant during the entire training for all training samples, and that is a suboptimal method since the recognition difficulty varies in training samples.
In additive angular margin softmax (AAM-Softmax) loss, the additional angular margin is set as a costant as well.
In this paper, we propose an APM-Softmax loss for language recognition with phoneitc multi-task learning, in which the additive phoneme-aware margin is automatically tuned for different training samples.
More specifically, the margin of language recognition is adjusted according to the results of phoneme recognition.
Experiments are reported on Oriental Language Recognition (OLR) datasets, and the proposed method improves AM-Softmax loss and AAM-Softmax loss in different language recognition testing conditions.

\end{abstract}

\section{Introduction}

Language recognition is a speech task to evaluate whether the target language is spoken in an auditory speech, and it is usually referred as Language Identification (LID), which is often set as the first process in a multilingual speech recognition system  \cite{5495082}.
To encourage the improvement of LID technologies and to tackle the real challenge existing in LID tasks, with the support of oriental language databases, the OLR Challenge has been organized annually since 2016, attracting dozens of teams around the world \cite{ap16,ap17,AP18-OLR,AP19-OLR,li2020ap20}.

LID and Automatic Speaker Verification (ASV) are two related speech tasks, sharing some key techniques.
The Gaussian Mixture Model (GMM) based generative models with back-end processing have been dominant over the past decades \cite{Dehak2011Front, languagerecog, langiv}.
In recent years, researchers have focused on exploring the usage of  Deep Neural Network (DNN) in LID to improve performance in various testing conditions \cite{2014Automatic}.
There were still many DNN based studies using i-vector, such as DNN i-vector \cite{6853887}, Bottleneck Feature (BNF) i-vector \cite{7080838}.
Later, end-to-end systems were also used to recognize languages \cite{6854622}.
More recently, motivated by the success of x-vector \cite{Snyder2018Spoken} in ASV, the DNN based language embedding has become a popular approach \cite{dnnlang}.
Based on the traditional x-vector structure, some deeper neural network structures were proposed, with the goal of learning a stronger language embedding extractor, such as CNN based x-vector \cite{06}, extended TDNN based x-vector \cite{07} and TDNN-F based x-vector \cite{08}.

The usage of phonetic information for ASV and LID tasks has also been studied.
There is a variety of learning strategies of phonetic information for ASV \cite{6853887,Kenny2014Deep, ontheusageofphonetic, inproceedingsmultitask, 9054333}.
However, there is still a lot of research space for the usage of phonetic information in the field of LID.
In \cite{Tian+2016}, the senone-based Long Short-Term Memory (LSTM) Recurrent Neural Network (RNN) was investigated to model long-range correlations in speech signals.
In \cite{8070977}, a phonetic temporal model for LID was proposed.
In our previous work \cite{li2021deep}, the phonetic multi-task learning framework was introduced into language recognition tasks.

Exploring a better loss function is always a hot topic in the community.
The goal of a loss function is to enlarge inter-class distance while intra-class distance is reduced.
Triplet loss \cite{li2017deep, zhang2017end} minimizes the distance between feature pairs from the same classes and maximizes the distance between feature pairs from the different classes.
Center loss was proposed in \cite{li2018deep,yadav2018learning}.
Meanwhile, efforts have also been made to improve the Softmax loss with angular space, such as Angular Softmax (A-Softmax), Additive Margin Softmax (AM-Softmax) loss, and Additive Angular Margin Softmax (AAM-Softmax) loss \cite{liu2017sphereface,li2018angular,wang2018additive,hajibabaei2018unified,yu2019ensemble,liu2019large,deng2019arcface}.

In AM-Softmax loss \cite{wang2018additive}, the additive margin is a hyper-parameter, which is tuned before training and is a fixed constant for all training samples during the entire training period.
While the angle between the feature vector and the center of its ground-truth class is scarcely the same for all training samples and the additive margin requires adaptive changes during training.
And the additive angular margin is also a fixed hyper-parameter in AAM-Softmax loss \cite{deng2019arcface}.
Further more, the usage of phonetic information in language-phoneme multi-task learning is inadequate, since it is only used as an auxiliary term in the final loss.
Thus, it is more reasonable to use a dynamic margin or a dynamic angular margin based on phonetic information to improve AM-Softmax loss or AAM-Softmax loss.
In this paper, we present an APM-Softmax for assisting language-phoneme multi-task learning.
Specifically, the additive phoneme-aware margin for language classification is adjusted according to the cumulative highest posterior probability from phoneme samples of a segment speech.
Experimental results on OLR datasets \cite{li2020ap20} show the effectiveness of the proposed APM-Softmax loss.

The rest of this paper is organized as follows.
Section 2 introduces the related Softmax, A-Softmax, AM-Softmax, and AAM-Softmax loss.
Section 3 presents the proposed additive phoneme-aware margin softmax loss.
Section 4 shows the experimental setup and Section 5 gives experimental results and analysis.
Finally, Section 6 concludes this paper.

\section{Related Works}

\subsection{Margin Softmax Loss}

The conventional Softmax loss is presented as:

\begin{equation}
L_{S}=-\frac{1}{N}\sum_{i=1}^{N}\log\frac{e^{W_{y_{i}}^{T}x_{i}+b_{y_{i}}}}{\sum _{j=1}^{C}e^{W_{j}^{T}x_{i}+b_{j}}}
\end{equation}

\noindent where $N$ is the number of training samples in a mini batch, $C$ indicates the number of languages in the training set, $x_{i}$ is the feature representation from the output of the last hidden layer of the $i$-th sample, $y_{i}$ is the ground truth label of the $i$-th sample, $W_{j}$ is the $j$-th column of the weights in the output layer, and $b_{j}$ indicates the bias term.

The logit  $W_{y_{i}}^{T}x_{i}+b_{y_{i}}$ can also be transformed to :

\begin{equation}
\left \| W_{y_{i}} \right \|\left \| x_{i} \right \|\cos\left ( \Theta _{y_{i}} \right )+b_{y_{i}}
\end{equation}

\noindent  where $\Theta _{y_{i}}$ is the angle between $W_{y_{i}}$ and $x_{i}$.

In A-Softmax \cite{li2018angular}, the weight vector is normalized and the bias term is discarded, in which $\left \| W_{y_{i}} \right \| = 1$ and $b_{y_{i}} = 0$, and the loss is presented as:

\begin{equation}
L_{AS}=-\frac{1}{N}\sum_{N}^{i=1}\log\frac{e^{\left \| x_{i} \right \|\phi \left ( \Theta _{y_{i}} \right )}}{e^{\left \| x_{i} \right \|\phi \left ( \Theta _{y_{i}} \right )}+\sum_{j=1;j\neq y_{i}}^{C}e^{\left \| x_{i} \right \|\cos\left ( \Theta _{j} \right )}}
\end{equation}

In the above equation, the original cosine angle $\cos\left ( \Theta _{y_{i}} \right )$ is replaced with $\phi \left ( \Theta _{y_{i}} \right )$:

\begin{equation}
\phi \left ( \Theta _{y_{i}} \right )=\left ( -1 \right )^{k}\cos\left ( m\Theta _{y_{i}} \right )-2k
\end{equation}

\noindent  where $\Theta _{y_{i}}\in \left [ \frac{k\pi }{m} ,\frac{\left ( k+1 \right )\pi }{m}\right ]$ and $k\in \left [ 0,m-1 \right ]$. The parameter $m$ is used to control the size of the angular margin, which is a positive integer.

The AM-Softmax loss \cite{wang2018additive} further extends the angular based Softmax losses, which is written as:

\begin{equation}
L_{AMS}=-\frac{1}{N}\sum_{i=1}^{N}\log\frac{e^{s\cdot \phi \left ( \Theta _{y_{i}} \right )}}{e^{s\cdot \phi \left ( \Theta _{y_{i}} \right )  }+\sum_{j=1;j\neq y_{i}}^{C}e^{s\cdot \cos\left ( \Theta _{j} \right )}}
\end{equation}

\begin{equation}
\phi \left ( \Theta _{y_{i}} \right )=\cos\left ( \Theta _{y_{i}} \right )-m
\end{equation}

\noindent where the feature vector $x_{i}$ in (3) is normalized to the unit, and is replaced with a hyper-parameter $s$.

The margin is larger with an additive penalty term $m$, which is also a hyper-parameter and is usually set as larger than zero.

More recently, AAM-Softmax loss \cite{deng2019arcface} improves AM-Softmax \cite{wang2018additive} by adding the penalty on the angle directly, then Equation (5) and (6) are rewritten as:

\begin{equation}
L_{AAMS}=-\frac{1}{N}\sum_{i=1}^{N}\log\frac{e^{s\cdot \phi \left ( \Theta _{y_{i}} \right )}}{e^{s \cdot \phi \left ( \Theta _{y_{i}} \right )}+\sum_{j=1;j\neq y_{i}}^{C}e^{s\cdot \cos\Theta _{j}}}
\end{equation}

\begin{equation}
\phi \left ( \Theta _{y_{i}} \right )=\cos\left ( \Theta _{y_{i}} +m \right )
\end{equation}

\noindent where the angle $\Theta _{y_{i}}$ is directly penal by an additive penalty term $m$, which is a pre-set hyper-parameter as well.

\subsection{Multi-task Learning}

\begin{figure}[t]
	\setlength{\belowcaptionskip}{0pt}
	\centering
	\includegraphics[width=220pt]{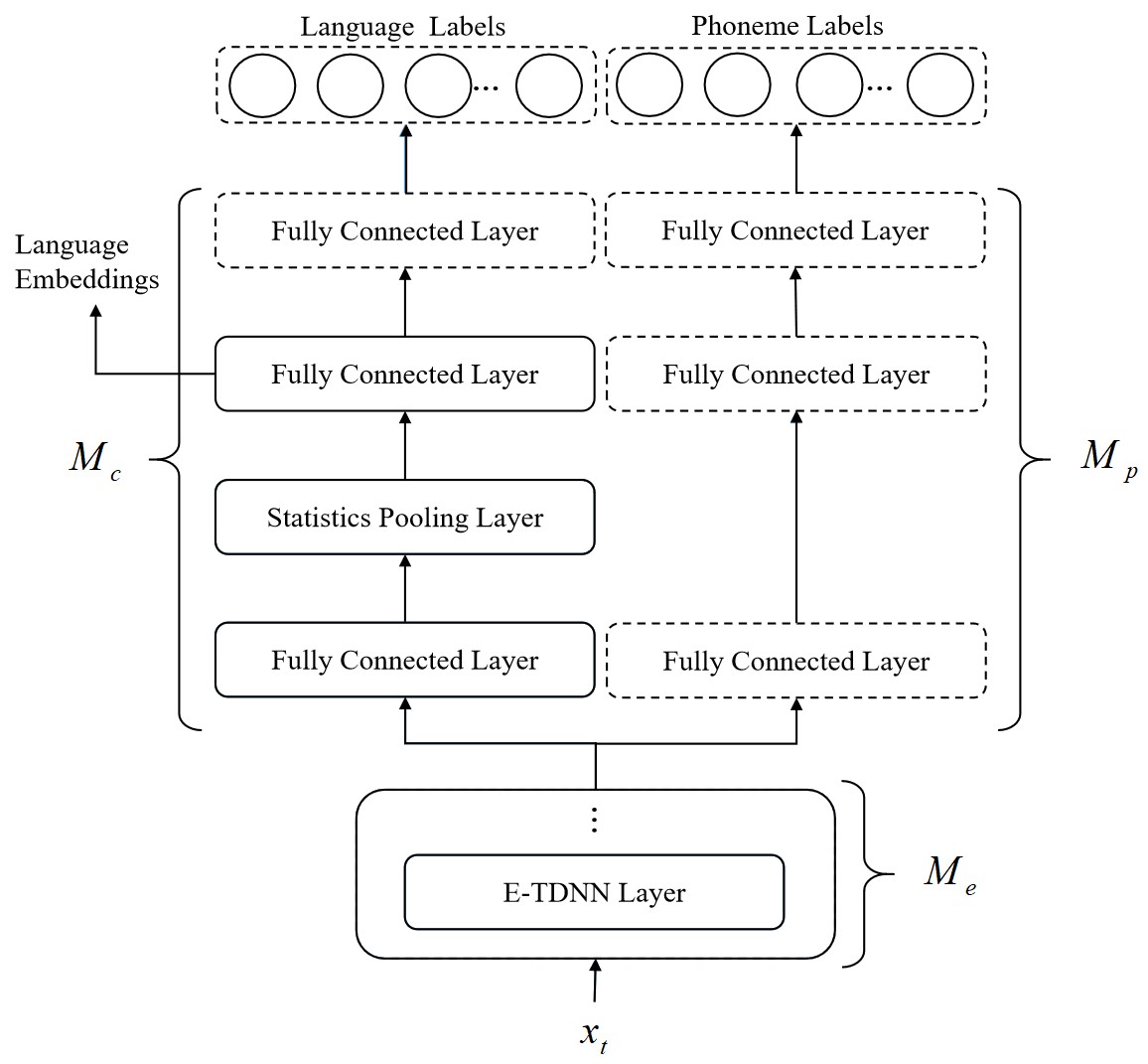}
	\caption{The language-phoneme multi-task learning structure. The dashed rectangles are removed after training.}
	\label{The multi-task learning structure}
\end{figure}

In this paper, we study the language-phoneme multi-task learning architecture \cite{ontheusageofphonetic,inproceedingsmultitask} based on the Extended Time Delay Nerual Network (E-TDNN) \cite{tdnne}, as shown in Figure \ref{The multi-task learning structure}.
The language-phoneme multi-task learning framework contains three modules, including the shared E-TDNN feature encoder $M_e$ at the frame-level, the phoneme classifier $M_p$ at the frame-level and the language classifier $M_c$ at the segment-level.
Given a segment $S$ of $T$ frames $X = \{x_{1},...,x_{t},...,x_{T}\}$, the total loss of multi-task learning is composed of the language classification loss $L_{c}$ and the phoneme classification loss $L_{p}$ with an empirical control factor $\alpha $, written as:

\begin{equation}
L_{total}=L_{c}+\alpha \cdot L_{p}
\end{equation}

\begin{equation}
L_{c}=CE\left ( M_{c}\left ( M_{e}\left ( X \right ) \right ),y^{S} \right )
\end{equation}

\begin{equation}
L_{p}=\frac{1}{T}\sum_{t=1}^{T}CE\left ( M_{p}\left ( M_{e}\left ( x_{t} \right ) \right ),y^{p} \right )
\end{equation}

\noindent where CE$(A, B)$ indicates the Coss Entropy loss (Softmax loss) computed between the two distributions $A$ and $B$. $y^{S}$ denotes the segment-level language label, and $y^{p}$ is the frame-level phoneme label.

After training, the dashed rectangles in Figure \ref{The multi-task learning structure} are removed from the model, and the language embeddings are extracted from the penultimate layer in the language classifier $M_{c}$.

\section{Additive Phoneme-aware Margin Softmax Loss}


As mentioned in Section 2.1, in AM-Softmax loss or AAM-Softmax loss, the additive margin or additive angular margin is a constant during the entire training period for all training samples.
However, the classification difficulty is barely the same among training samples, and it changes during the training.
Thus, it is better to use a dynamic additive margin according the training samples rather than using a fixed one.
Further, the phonetic information is not fully utilized in language-phoneme multi-task learning mentioned in Section 2.2, since the phoneme classification sub-network is only used to constitute an auxiliary term in the total loss function.

\begin{figure}[h]
	\setlength{\belowcaptionskip}{0pt}
	\centering
	\includegraphics[width=220pt]{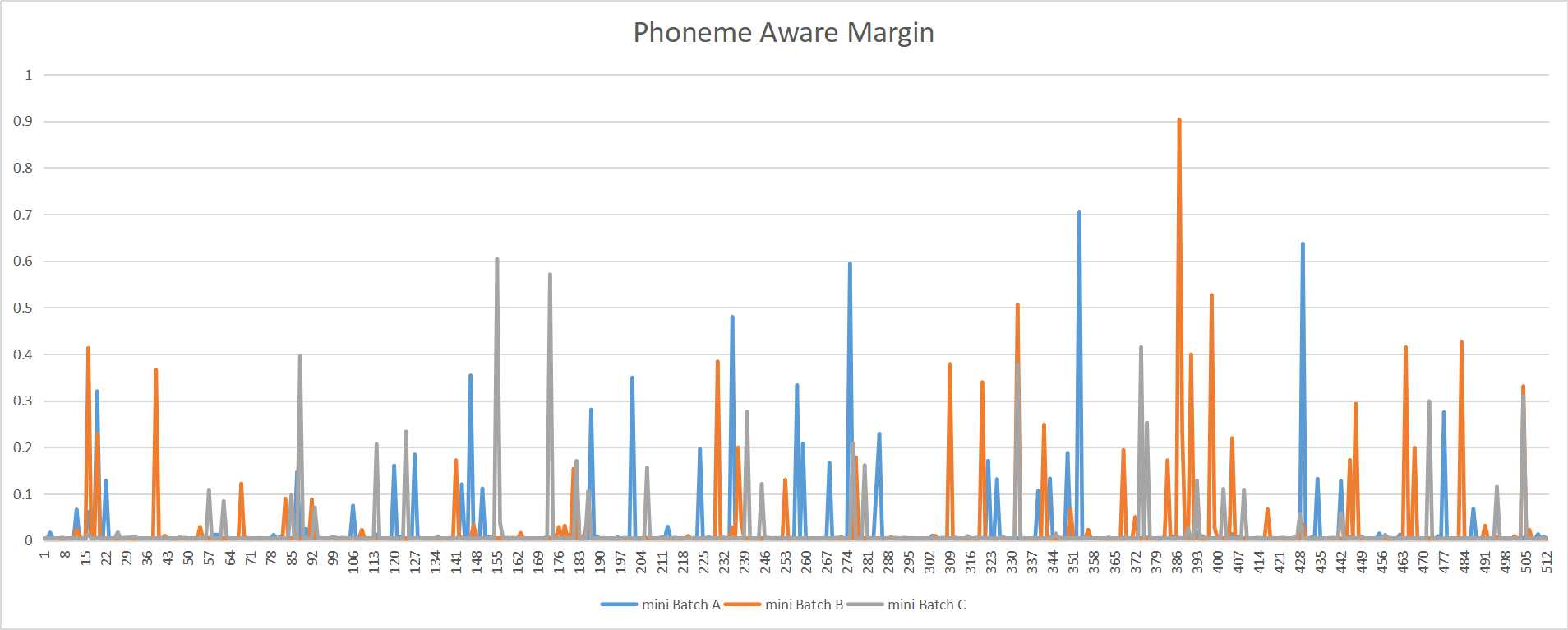}
	\caption{The additive phoneme-aware margins $ \beta \cdot p$ in three random-selected mini batches, with the control factor $\beta  = 10$. Each color indicates training samples from a mini batch.}
	\label{phoneme aware margin}
\end{figure}

This paper proposes an APM-Softmax loss for assisting language-phoneme multi-task learning, replacing the conventional Softmax loss in Equation (10).
In brief, this approach introduces an adaptive phoneme-aware margin according to the cumulative highest posterior probabilities from phoneme samples in a speech segment.
The proposed adaptive phoneme-aware margin is defined as:

\begin{equation}
P_{i}=m+ \beta  \cdot p_{i}
\end{equation}

\noindent where $m$ is a fixed basic additive (angular) margin, $\beta $ is a control factor to control the strength of the phoneme-aware margin.

\begin{equation}
p_{i}= \frac{1}{T} \sum_{t=1}^{T}\max\left ( {\rm Softmax}\left ( M_{p}\left ( M_{e}\left ( x_{t,i} \right ) \right ) \right ) \right )
\end{equation}

\noindent where $T$ is the length of a chunk size in the training sample, $\max({\rm Softmax}(\cdot))$ is the operation for selecting the maximal posterior probability from the phoneme recognition sub-network's computation $M_{p}\left ( M_{e}\left ( \cdot \right )\right )$ of the traing feature $x_{t,i}$.

Introducing the proposed adaptive phoneme-aware margin into AM-Softmax loss, the APM-Softmax is rewritten as:

\begin{equation}
L_{APMS}=-\frac{1}{N}\sum_{i=1}^{N}\log\frac{e^{s\cdot \rho  \left ( \Theta _{y_{i}} \right )}}{e^{s\cdot \rho  \left ( \Theta _{y_{i}} \right )}+\sum_{j=1;j\neq y_{i}}^{C}e^{s\cdot \cos\left ( \Theta _{j} \right )}}
\end{equation}

\begin{equation}
\rho  \left ( \Theta _{y_{i}} \right ) =\cos\left ( \Theta _{y_{i}} \right ) - P_{i}
\end{equation}

\noindent where the angular margin $\cos\left ( \Theta _{y_{i}} \right )$ is affected by the additive phoneme-aware margin $P_{i}$, which is changing among training samples $i$.

And into AAM-Softmax loss, Equation (15) is rewritten as:

\begin{equation}
\rho  \left ( \Theta _{y_{i}} \right ) =\cos\left ( \Theta _{y_{i}} + P_{i} \right )
\end{equation}

\noindent where the angle $\Theta _{y_{i}}$ is directly impacted by the dynamic penalty term $P_{i}$.

As shown in Equation (13), in APM-Softmax, we use the average maximal phoneme posterior probability in a chunk of $T$ features to indicate the degree of utterance level phoneme-aware, and it varies from training samples in every mini batch.
It should be noted that, only the highest posterior probability is used for indicating the recognition difficulty of phonemes, rather than the posterior probability corresponding to its ground truth label.

Figure \ref{phoneme aware margin} shows the additive phoneme-aware margins $ \beta \cdot p$ in three random-selected mini batches, with the control factor $\beta  = 10$.
In Figure \ref{phoneme aware margin}, it can be observed that in each mini batch, the number of noteworthy additive phoneme-aware margins are minor.
The number of those peaks are similar among different mini batches, indicating that training samples with high posterior probabilities  are distributed evenly among mini batches, which is significant for the stationary iteration of training.

\begin{figure}[]
	\setlength{\belowcaptionskip}{-7pt}
	\centering
	\includegraphics[width=230pt]{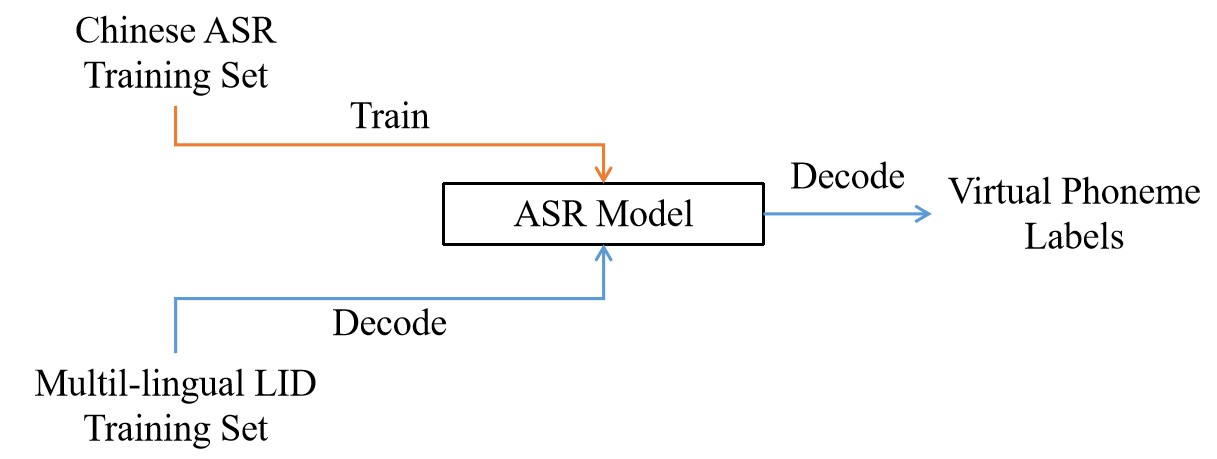}
	\caption{Preparation of virtual phoneme labels for language-phoneme multi-task learning.}
	\label{phone}
\end{figure}

\section{Experimental Setup}

\subsection{Datasets}
The language recognition database used in our experiments is the OLR dataset \cite{ap16,ap17,AP18-OLR,AP19-OLR,li2020ap20} for both training and testing.

In the OLR challenges, additional training materials were prohibited and participants were only permitted to use several specified datasets, including AP16-OL7, AP17-OL3, AP17-OLR-test, AP18-OLR-test, and THCHS-30 (plus the accompanying resources, such as dictionary and transcriptions) \cite{THCHS30_2015}.

To better investigate of the proposed method, three kinds of test sets were used, which correspond to three particular LID conditions, including short utterance LID, cross-channel LID, and open-set dialect identification.

For short utterance LID, the AP18-OLR-test-short was used as the test set, with about 5.8 hours of speech. This data set is a close-set identification set, in which the languages of the utterances are as in the training set, but the test utterances are only one second long.
For cross-channel LID, the AP19-OLR-test-channel was used, with about 8.9 hours of speech. This test data is recorded in different channels and is also a close-set identification set, including six known target languages: Cantonese, Indonesian, Japanese, Russian, Korean, and Vietnamese.
For open-set dialect identification, the AP20-OLR-dev-task2 was used, with about 10.21 hours of utterances in total. The AP20-OLR-dev-task2 contains the dialect test subset of AP19-OLR-dev-task3, which contains three target dialects: Hokkien, Sichuanese, and Shanghainese, and the test subset of AP19-OLR-test-task3, which contains three nontarget (interfering) languages: Catalan, Greek, and Telugu.

\begin{table*}[ht]
			\setlength{\belowcaptionskip}{0pt}
					\centering
		\caption{ Experimental results on OLR dataset with the metric value $C_{avg}$.}
	\begin{tabular}{|c|c|c|c|c|c|c|c|c|}
		\hline
		\multirow{2}{*}{No.} & \multirow{2}{*}{System}      & \multirow{2}{*}{Loss} & \multicolumn{3}{c|}{Margin (P)}   & \multicolumn{3}{c|}{Test Sets (Cavg)}         \\ \cline{4-9}
		&                              &                       & $m$        & $\beta$      & $p$          & Short-utt & Cross-channel   & Open-set        \\ \hline
		1                    & x-vector                     & \multirow{2}{*}{$L_{S}$}    & \multicolumn{3}{c|}{-}            & 0.0540    & 0.2694          & 0.0851          \\ \cline{1-2} \cline{4-9}
		2                    & \multirow{7}{*}{MT x-vector} &                       & \multicolumn{3}{c|}{-}            & 0.0470    & 0.2779          & 0.0735          \\ \cline{1-1} \cline{3-9}
		3                    &                              & \multirow{2}{*}{$L_{AMS}$}  & 0.2      & \multicolumn{2}{c|}{-} & 0.0458    & 0.2769          & 0.1032          \\ \cline{1-1} \cline{4-9}
		4                    &                              &                       & 0.02     & \multicolumn{2}{c|}{-} & 0.0467    & 0.2458          & 0.0792          \\ \cline{1-1} \cline{3-9}
		5                    &                              & $L_{APMS}$                  & 0.2      & 10        & 0.002      & 0.0463    & \textbf{0.2497} & \textbf{0.0707} \\ \cline{1-1} \cline{3-9}
		6                    &                              & \multirow{2}{*}{$L_{AAMS}$} & 0.2      & \multicolumn{2}{c|}{-} & 0.0451    & 0.2731          & 0.0941          \\ \cline{1-1} \cline{4-9}
		7                    &                              &                       & 0.02     & \multicolumn{2}{c|}{-} & 0.0441    & 0.274           & 0.0874          \\ \cline{1-1} \cline{3-9}
		8                    &                              & $L_{APAMS}$                 & 0.2      & 10        & 0.002      & 0.0445    & 0.2590          & 0.0761          \\ \hline
	\end{tabular}
\end{table*}

\subsection{Phoneme Labels}

For training a language-phoneme multi-task learning model, phoneme labels for the multi-lingual LID training set are necessary.
Due to the constraint of the OLR Challenge rules, we only used THCHS-30 to get an ASR model, and to generate phoneme labels for the multi-lingual LID traning set, as shown in Figure  \ref{phone}.
Therefore, every language in the LID training set was decoded corresponding to Chinese phonemes, and this can be seem as virtual phonemes.

THCHS-30 \cite{THCHS30_2015} is an allowed-to-use data set in the OLR Challenges, which contains the lexicon and dictionary files for 30 hours of Chinese ASR training speech.
Thus, based on the THCHS-30 training data, according to the Kaldi's recipe \cite{kaldi}, a $tri4b$ (tri-phone) GMM-based ASR model was trained, including 1,507 Probability Density Function Identifications (PDF-IDs).
For the purpose of accurate phoneme alignment, we used the PDF-IDs in each frame to represent the corresponding phoneme labels.
Finally, the virtual phoneme labels for the multi-lingual OLR training set were decoded from the trained Chinese ASR model.

\subsection{Training Details}
Before training, we carried out data augmentation, including speed and volume perturbation, to increase the amount and diversity of the training data.
The language-phoneme multi-task learning network was developed based on the OLR 2020 baseline E-TDNN x-vector, and we only added the phoneme classifier $M_{p}$ as mention in Section 2.2.
The acoustic features is 20-dimensional MFCC with 3-dimensional pitch, which has frame-length of 25ms, frame shifts of 10ms, and mean normalization over a sliding window of up to three seconds.
Voice Activity Detection (VAD) was used to filter out non-speech frames.
The back-end process was the same for all three test sets when the embeddings were extracted, as mentioned in the baseline system on OLR 2020 Challenge \cite{li2020ap20}.
The front-end models were based on PyTorch \cite{paszke2017automatic,Tong2021subtools} optimized with Adam optimizer, with a mini-batch size of 512, and a chunk size of 100 frames in each segment, while the acoustic feature extraction and the back-end process were executed on Kaldi \cite{kaldi}.
The hyper-parameter $\beta$ of the phoneme-aware margin is empirically set to 10.

\section{Results and Analysis}

In this paper, the primary metric $C_{avg}$ in \cite{li2020ap20} is chosen to evaluate the language recognition systems.
Experimental results are given in Table 1.
Thereinto, Softmax loss, AM-Softmax loss and AAM-Softmax loss are abbreviated as $L_{S}$, $L_{AMS}$ and $L_{AAMS}$ respectively.
The AM-Softmax based APM-Softmax loss in Equation (15) is named as $L_{APMS}$, and AAM-Softmax based APM-Softmax loss in Equation (16) is as $L_{APAMS}$.
In Table 1, x-vector indicates the baseline E-TDNN x-vector system and MT x-vector is for the multi-task learning E-TDNN x-vector system mentionded in 2.2.

Compared with the baseline x-vector in system No. 1, the improvement provided by language-phoneme multi-task learning is reported in system No. 2.
Although the phoneme labels used to train the language-phoneme multi-task learning model were from a Chinese ASR model, which are named virtual phoneme labels in this paper, the introduction of phonetic information significantly improves the language recognition performance in Short-uttrance task and Open-set task.

Moreover, it can be observed that introducing larger margins into language-phoneme multi-task learning x-vector models, as shown in system No. 3 and No. 6, improves the performance in Short-utterance task and Cross-channel task, while degrades that in Open-set task.

In order to further verify the effectiveness of the proposed APM-Softmax losses, based on the language-phoneme multi-task learning network, Table 1 shows the comparison of AM-Softmax loss and AAM-Softmax loss with a little bit larger $m$, for compensating the additive phoneme margin $\beta \cdot p$ introduced into APM-Softmax losses.
As shown in Figure \ref{phoneme aware margin}, the additive phoneme margin $p$ is varying within each mini batch during the training, thus, the $p$ listed in Table 1 is the mean value from training samples of all mini batches, with $\beta  = 10$.
With a larger $m$, as reported in system No. 5, the performance of AM-Softmax loss is slightly improved in Cross-channel and Open-set tasks, but there is still a significant performance gap compared with APM-Softmax loss, in three LID tasks.
And in AAM-Softmax, as in system No. 7, with a larger $m$, better results are achieved in Short-uttreance task and Open-set task, while its contribution in those two tasks is much less than the proposed APM-Softmax, as in system No. 8.
Moreover, the degradation on the Open-set task introduced by larger margin training strategies are significantly sloved with the dynamic additive phoneme-aware margin.

The results reinforce the assumption that dynamically setting the margin or angular margin according to the phonetic information for each training sample is more reasonable than using a constant margin or angular margin shared by all training samples.

\section{Conclusions}

In this paper, we propose an APM-Softmax loss for improving language-phoneme multi-task learning.
In the presented APM-Softmax loss, the additive margin changes according to the phoneme recognition results of each langauge training samples during training, which improves the fixed margin AM-Softmax loss, and the fixed angular margin AAM-Softmax loss.
The phoneme-aware method adopts an adaptive margin according to the phonetic information.
We investigated the proposed method with the E-TDNN based multi-task learning architecture on OLR dataset.
Experimental results show that our proposal improves the AM-Softmax loss and the AAM-Softmax loss based systems, under three different language recognition conditions.

In the future work, more usages of the phonetic information for language recognition will be investigated.
Moreover, the proposed additive phoneme-aware margin loss is also suitable for other speech classification tasks, such as speaker recognition task, thus, more studies will be given in the applications of APM-Softmax loss.

\section{Acknowledgements}

This work is supported by the National Natural Science Foundation of China (Grant No. 61876160, Grant No. 62001405).

\bibliographystyle{IEEEtran}

\bibliography{mybibfile}

\end{document}